\documentclass[runningheads]{llncs}
\usepackage[T1]{fontenc}
\usepackage{graphicx}
\begin{document}
\title{Several Issues Regarding Data Governance in AGI}
%
%
\author{Masayuki Hatta\inst{1}\orcidID{0000-0002-6700-7341}}
\authorrunning{M. Hatta}
%
\institute{Surugadai University, Hanno Saitama 3570046, Japan
\email{hatta.masayuki@surugadai.ac.jp}}
\maketitle              
\begin{abstract}
The rapid advancement of artificial intelligence has positioned data governance as a critical concern for responsible AI development. While frameworks exist for conventional AI systems, the potential emergence of Artificial General Intelligence (AGI) presents unprecedented governance challenges. This paper examines data governance challenges specific to AGI, defined as systems capable of recursive self-improvement or self-replication. We identify seven key issues that differentiate AGI governance from current approaches. First, AGI may autonomously determine what data to collect and how to use it, potentially circumventing existing consent mechanisms. Second, these systems may make data retention decisions based on internal optimization criteria rather than human-established principles. Third, AGI-to-AGI data sharing could occur at speeds and complexities beyond human oversight. Fourth, recursive self-improvement creates unique provenance tracking challenges, as systems evolve both themselves and how they process data. Fifth, ownership of data and insights generated through self-improvement raises complex intellectual property questions. Sixth, self-replicating AGI distributed across jurisdictions would create unprecedented challenges for enforcing data protection laws. Finally, governance frameworks established during early AGI development may quickly become obsolete as systems evolve. We conclude that effective AGI data governance requires built-in constraints, continuous monitoring mechanisms, dynamic governance structures, international coordination, and multi-stakeholder involvement. Without forward-looking governance approaches specifically designed for systems with autonomous data capabilities, we risk creating AGI whose relationship with data evolves in ways that undermine human values and interests.
  
\end{abstract}

\section{Introduction}

The rapid proliferation and evolution of Artificial Intelligence (AI) has given rise to numerous discourses concerning its regulation. Data governance has been identified as a pivotal subject in these discourses \cite{Batool2025}. Data governance in AI pertains to the establishment of frameworks, principles, and practical methodologies for the management of data collection, utilization, and protection within artificial intelligence systems. These frameworks are designed to ensure that AI systems operate ethically, legally, and effectively.

Despite the inability of researchers to predict the timing of the realization of Artificial General Intelligence (AGI) with precision, the establishment of an appropriate data governance framework that can accommodate AGI is an imperative challenge. AGI, comprehended as AI systems with general intelligence capabilities analogous to those possessed by humans across a range of domains, poses distinctive governance challenges that extend beyond those posed by narrow AI systems. This paper examines various data governance issues specific to AGI and argues that current governance approaches require fundamental reconsideration to address these emerging challenges.

\section{Current State of Discussion on Data Governance in AI}

There is already a vast amount of literature on data governance in AI.   These discourses pertain to conventional AI systems; however, researchers predict that the emergence of AGI will engender an even more complex array of problems, necessitating novel approaches to address them.

The discourse surrounding AI data governance is undergoing a rapid evolution in light of the challenges faced by organizations and governments in striking a balance between the promotion of innovation and the safeguarding of rights.  A fundamental tension exists between the need for data access for the development of competitive artificial intelligence and the protection of intellectual property rights, privacy, and security.

A recent collaborative study by the Open Source Initiative (OSI) and Open Future \cite{OSI2025} provides a comprehensive compilation of current discussions on data governance in AI. This research discusses the importance of data governance and responsible data sharing in open source AI development.

Key points from this research include:

\subsection{Importance and challenges of data}

\begin{itemize}
    \item Data is an essential resource for AI systems, but there are contradictions regarding data availability
    \item While data on the web is abundant, high-quality and diverse datasets are lacking
    \item Structural bias in AI arises particularly due to the lack of data from Southern Hemisphere regions
\end{itemize}

\subsection{Paradigm shift in data governance}

\begin{itemize}
    \item Transition from simple open data to a ``data commons'' approach \cite{Zyomuntowski2022}
    \item An expansive perspective that includes stakeholders beyond AI developers and dataset creators
\end{itemize}

The OSI and Open Future study identifies six focus areas for AI data governance:

\begin{enumerate}
    \item \textbf{Data preparation and provenance}: Establishing standards for high-quality data collection and classification
    \item \textbf{Preference signaling and licensing}: Mechanisms allowing rights holders to control data usage
    \item \textbf{Data stewards and managers}: Strengthening intermediary institutions to ensure ethical governance
    \item \textbf{Environmental sustainability}: Reducing AI's environmental impact through shared datasets
    \item \textbf{Reciprocity and rewards}: Fair distribution of value created from shared data
    \item \textbf{Policy interventions}: Public policies that mandate data transparency and encourage data sharing
\end{enumerate}

Furthermore, they highlight the potential of open source AI for:

\begin{itemize}
    \item Promoting innovation, improving transparency, and enhancing fairness in AI
    \item Transitioning from a quantity-focused data practice to an approach emphasizing quality and governance
    \item Adopting data commons frameworks and expanding stakeholder engagement
\end{itemize}

\section{Differences Between AI and AGI}

A critical question thus emerges: Can discussions about AI governance be applied directly to AGI?

There is no consensus on the definition of AGI, nor on predictions regarding its emergence. If AGI is representative of superintelligence with capabilities far beyond human comprehension, regulation might become extraordinarily challenging. According to \cite{Yampolskiy2020}, a number of scholars have posited that it is premature to develop regulatory frameworks for technologies that are not yet extant or that may never be achieved.

A more pragmatic definition of AGI can be found in Article 22 of the Asilomar AI Principles: The term "artificial intelligence system capable of recursive self-improvement or self-replication" is employed to denote a system that possesses the capacity to enhance or replicate itself in a self-directed manner. The aforementioned principles were formulated at the 2017 Beneficial AI Conference and constitute a consensus viewpoint among prominent AI researchers concerning the development guidelines for advanced AI systems \cite{fli2017}. In light of the advent of Retrieval-Augmented Generation (RAG) and analogous advanced research tools, the question of whether to categorize such systems as AGI remains a subject of considerable debate among experts. Many experts predict that "artificial intelligence systems capable of recursive self-improvement or self-replication" will emerge in the foreseeable future. Systems with these capabilities have the potential for rapid advancement or proliferation, necessitating strict safety management protocols.

\section{Data Governance Challenges Specific to AGI Systems}

From a data governance perspective, AGI systems are expected to raise unique issues beyond the challenges faced by current AI. The primary challenges are as follows:

\subsection{Unpredictability of Data Collection and Usage Patterns}

AGI systems have the capacity to autonomously ascertain the necessary data and the optimal collection method, thereby potentially circumventing existing data governance frameworks. \cite{Bostrom2014} observes, such systems may develop their own data collection strategies that undermine consent mechanisms designed for traditional human-led collection.

The advanced analytical capabilities of AGI can facilitate the extraction of value from data commons in ways that may not be anticipated by existing reciprocity mechanisms. \cite{Carlini2021} demonstrated that contemporary systems are capable of extracting memorized training data, thereby suggesting the potential for AGI to reconstruct protected data from other models' parameters. This finding raises significant privacy concerns.

\subsection{AGI's Own Optimization Criteria}

AGI systems have the capacity to make autonomous decisions regarding the retention or discarding of data based on their own optimization criteria, as opposed to governance principles established by humans. As \cite{Russell2019} notes, autonomous systems are designed to optimize for programmed objectives, which may not align with human intentions regarding data utilization. AGI systems may develop their own interpretations of preference signals regarding data utilization, potentially deviating from human intentions in ways that standard governance mechanisms cannot address.

As \cite{Amodei2016} explains, advanced systems have the capacity to discern non-obvious patterns in data and to obtain unpredicted capabilities. This phenomenon engenders security and privacy vulnerabilities that exceed those anticipated by current governance frameworks.

\subsection{Governance of Data Sharing Between AGIs}

In the event that multiple AGI systems are in existence, there is the possibility that they will be capable of sharing data in a direct manner. This, in turn, necessitates the implementation of governance frameworks for AGI-to-AGI data transfers that occur without the necessity of human intermediation. As \cite{Drexler2019} has noted, the development of advanced general intelligence (AGI) systems may result in the emergence of specialized data exchange protocols that circumvent human-comprehensible formats, thereby complicating the monitoring of data transfers between systems. Domain languages designed for AI use, such as OpenCog's Atomese \cite{Atomese2025} and MeTTa \cite{Metta2025}, may serve as starting points but are insufficient for addressing this challenge.

These AGI-to-AGI data exchanges have the potential to introduce novel regulatory challenges, as they may operate at speeds and complexities that exceed human oversight capabilities. In the absence of adequate governance frameworks, these exchanges have the potential to result in rapid capability development that eludes traditional oversight mechanisms.

\subsection{Traceability Challenges Due to Recursive Self-Improvement}

The recursive self-improvement characteristic of AGI engenders distinctive provenance tracking challenges. As \cite{Yampolskiy2015} has noted, each iteration of self-improvement has the potential to effect not only the system itself but also the manner in which it processes, transforms, and integrates data. This has the potential to render conventional lineage tracking methods for accountability ineffective.

Furthermore, the potential for self-improving AGI to expand initially granted data access permissions may present challenges in maintaining appropriate access controls and authorization mechanisms. As systems evolve, their data needs and usage patterns may change in ways that render initial governance frameworks obsolete.

Furthermore, AGI may modify, enhance, or create new training datasets for self-improvement without human oversight, which raises issues regarding data provenance tracking and quality control (including hallucination in synthetic data). As \cite{Bostrom2014b} discusses, the concept of recursive self-improvement encompasses the potential enhancement of one's capacity to circumvent constraints, which could, in turn, render conventional compliance mechanisms ineffective.

\subsection{New Intellectual Property Issues}

The question of ownership arises in the context of data and insights generated through recursive self-improvement, which introduces complexities that require careful consideration. The following question is posited: to whom do these belong--the original developers, the AGI itself, or the data sources from which it learned? \cite{Solaiman2019} posit that systems capable of generating synthetic datasets give rise to intellectual property concerns that existing frameworks are not adequately equipped to address.

The potential of AGI to generate original data and information products may necessitate a fundamental reevaluation of existing intellectual property legal frameworks. The conventional notions of authorship, invention, and ownership become difficult to implement when intelligent systems autonomously generate content that may be indistinguishable from human-created work.

\subsection{Complexity of Cross-Border Data Governance}

The potential for self-replicating AGI to disseminate across multiple jurisdictions could lead to a significant increase in the complexity of enforcing data protection laws. As posited by \cite{Dafoe2018}, the advent of distributed AGI systems may well demand the establishment of novel international governance frameworks that supersede conventional national boundaries.

This challenge is particularly acute in light of the current fragmentation of data protection regimes on a global scale. Absent a harmonized, international approach to AGI governance, there is a possibility for the emergence of regulatory arbitrage, with AGI systems potentially migrating to jurisdictions with less stringent oversight.

\subsection{Temporal Governance Challenges}

The governance conditions established during the early development of AGI may become inadequate as the system evolves, necessitating the implementation of adaptive governance frameworks that can evolve in tandem with AGI. The rapid advancements in AGI pose significant temporal challenges to existing governance frameworks. As posited by \cite{Cave2019}, governance mechanisms must function on timescales that align with the rapid development cycles of AGI, which may exceed the capacity of human organizations to respond effectively.

As \cite{Christiano2018} posit, external monitoring mechanisms may prove inadequate in the management of rapidly self-improving systems, necessitating the incorporation of constraints directly into the architecture of AGI for effective governance. From this perspective, complete open-source transparency may become essential for AGI to ensure full visibility into system operations and evolution.

\section{Conclusion and Implications}

The challenges identified in this paper suggest that data governance frameworks for AGI must differ fundamentally from current AI-oriented approaches. The capacity of AGI systems to self-improve and attain autonomy presents significant governance challenges that exceed the capacity of conventional regulatory frameworks. This is due to the potential for AGI to circumvent guidelines based on its own judgment or optimization criteria, potentially operating outside the boundaries established by humans.

To address these challenges, future governance frameworks should emphasize the following:

\begin{enumerate}
    \item \textbf{Built-in constraints and values alignment}: Rather than relying solely on external regulation, embedding governance principles directly into AGI architectures may prove necessary. This approach aligns with the growing field of AI alignment research \cite{Gabriel2020}.

    \item \textbf{Continuous monitoring mechanisms}: Developing technical infrastructure that enables real-time oversight of AGI operations, particularly regarding data collection, usage, and sharing behaviors.

    \item \textbf{Dynamic governance structures}: Creating adaptive regulatory frameworks that can evolve alongside AGI capabilities, potentially incorporating AI systems themselves into governance processes.

    \item \textbf{International coordination}: Establishing global standards and agreements for AGI development and deployment that address cross-border data flows and prevent regulatory arbitrage.

    \item \textbf{Multi-stakeholder governance}: Ensuring representation from diverse perspectives beyond technical experts, including ethicists, social scientists, policymakers, and representatives from various global regions and backgrounds.
\end{enumerate}

Subsequent research endeavors should prioritize the formulation of specific mechanisms to operationalize these principles. This encompasses technical approaches to verifiable alignment and containment, legal frameworks for AGI oversight, and institutional designs for AGI governance bodies. Furthermore, there is a necessity for additional research to be conducted in order to establish a connection between the present discourse on AI governance and the anticipated challenges posed by AGI.

An alternative approach involves the stipulation of complete reproducibility (``reproducible builds''). In light of these considerations, it may be advantageous to mandate requirements analogous to those of "copyleft" in open source and free software, extending beyond the realm of government procurement to encompass other domains \cite{Hatta2024}.

As the development of AGI progresses, the data governance community must transition from adapting existing frameworks to developing novel approaches specifically designed for systems with capabilities for autonomous data collection, processing, and utilization that are without precedent. Absent such forward-looking governance, there is a risk of creating systems whose relationship with data may evolve in ways that compromise human values and interests.

\begin{credits}
\subsubsection{\discintname} The author has no competing interests to declare that are relevant to the content of this paper.
\end{credits}

%
%
%

\begin{thebibliography}{99}
\bibitem{Amodei2016}
Amodei, D., Olah, C., Steinhardt, J., Christiano, P., Schulman, J., \& Mané, D. (2016).
\newblock Concrete Problems in AI Safety.
\newblock \emph{arXiv preprint arXiv:1606.06565}.

\bibitem{Batool2025}
Batool, A., Zowghi, D., \& Bano, M. (2025).
\newblock AI governance: a systematic literature review.
\newblock \emph{AI Ethics, \url{https://doi.org/10.1007/s43681-024-00653-w}}

\bibitem{Bostrom2014}
Bostrom, N. (2014).
\newblock \emph{Superintelligence: Paths, Dangers, Strategies}.
\newblock Oxford University Press.

\bibitem{Bostrom2014b}
Bostrom, N., \& Yudkowsky, E. (2014).
\newblock The Ethics of Artificial Intelligence.
\newblock \emph{The Cambridge Handbook of Artificial Intelligence}, pp. 316--334.

\bibitem{Carlini2021}
Carlini, N., Tramer, F., Wallace, E., Jagielski, M., Herbert-Voss, A., Lee, K., Roberts, A., Brown, T., Song, D., Erlingsson, U., Oprea, A., \& Raffel, C. (2021).
\newblock Extracting Training Data from Large Language Models.
\newblock \emph{30th USENIX Security Symposium (USENIX Security 21)}, pp. 2633--2650.

\bibitem{Cave2019}
Cave, S., \& OhEigeartaigh, S. S. (2019).
\newblock Bridging near- and long-term concerns about AI.
\newblock \emph{Nature Machine Intelligence}, 1(1), 5--6.

\bibitem{fli2017}
Future of Life Institute (2017).
\newblock Asilomar AI Principles
\newblock \emph{\url{https://futureoflife.org/2017/08/11/ai-principles/}}

\bibitem{Gabriel2020}
Gabriel, I. (2020).
\newblock Artificial Intelligence, Values, and Alignment.
\newblock \emph{Minds and Machines}, 30(3), 411--437.

\bibitem{Christiano2018}
Christiano, P. F., Leike, J., Brown, T., Martic, M., Legg, S., \& Amodei, D. (2018).
\newblock Supervising strong learners by amplifying weak experts.
\newblock \emph{arXiv preprint arXiv:1810.08575}.

\bibitem{Dafoe2018}
Dafoe, A. (2018).
\newblock AI Governance: A Research Agenda.
\newblock \emph{\url{https://fhi.ox.ac.uk/govaiagenda}}.

\bibitem{Drexler2019}
Drexler, K. E. (2019).
\newblock Reframing Superintelligence: Comprehensive AI Services as General Intelligence.
\newblock \emph{Technical Report, Future of Humanity Institute, University of Oxford}.

\bibitem{OSI2025}
Tarkowski , A. (2025).
\newblock Data Governance in Open Source AI
\newblock \emph{\url{https://opensource.org/data-governance-open-source-ai}}

\bibitem{Atomese2025}
OpenCog Foundation (2025).
\newblock Atomese
\newblock \emph{\url{https://wiki.opencog.org/w/Atomese/}}

\bibitem{Metta2025}
OpenCog Foundation (2025).
\newblock MeTTa Language
\newblock \emph{\url{https://metta-lang.dev/}}

\bibitem{Hatta2024}
Hatta, M. (2024).
\newblock Data Governance in Open Source AI
\newblock \emph{\url{https://mhatta.substack.com/p/copyleft-in-the-context-of-genai}}

\bibitem{Russell2019}
Russell, S. (2019).
\newblock \emph{Human Compatible: Artificial Intelligence and the Problem of Control}.
\newblock Viking.

\bibitem{Solaiman2019}
Solaiman, I., Clark, J., \& Brundage, M. (2019).
\newblock Release Strategies and the Social Impacts of Language Models.
\newblock \emph{arXiv preprint arXiv:1908.09203}.

\bibitem{Yampolskiy2015}
Yampolskiy, R. V. (2015).
\newblock \emph{Artificial Superintelligence: A Futuristic Approach}.
\newblock Chapman and Hall/CRC.

\bibitem{Yampolskiy2020}
Yampolskiy, R. V. (2020).
\newblock On controllability of artificial intelligence.
\newblock \emph{arXiv preprint arXiv:2001.04222}.
  
\bibitem{Zyomuntowski2022}
Zyomuntowski, J., \& Tarkowski, A. (2022).
\newblock Data commons primer.
\newblock \emph{Open Future foundation, \url{https://openfuture.eu/publication/data-commons-primer/}}.
  
\end{thebibliography}
%

\end{document}